\documentclass[twoside]{ilcws10}
\usepackage[latin1]{inputenc}
\usepackage[dvips]{graphicx,epsfig,color}
\usepackage{wrapfig,rotating}
\usepackage{amssymb,amsmath,array}

\begin{document}
\title{
Comparison of hadron shower data with simulations} 
\author{Shaojun LU$^1$ 
\vspace{.3cm}\\
1- DESY FLC group \\
Notkestr. 85, Hamburg, 22607, Germany
}

\maketitle

\begin{abstract}
An analog hadron calorimeter (AHCAL) prototype of 5.3 nuclear interaction lengths
thickness has been designed and constructed by members of the CALICE Collaboration. The AHCAL prototype
consists of a 38-layer sandwich structure of steel plates and 7608 scintillator tiles that
are read out by wavelength-shifting fibres coupled to SiPMs. The signal is amplified and shaped
with a custom-designed ASIC. A calibration/monitoring system based on LED light was developed
to monitor the SiPM gain and to measure the full SiPM response curve in order to correct for non-linearity.
Ultimately, the physics goals are the study of hadronic shower shapes and testing the
concept of particle flow. The technical goal consists of measuring the performance and reliability
of 7608 SiPMs. The AHCAL prototype was commissioned in test beams at DESY, CERN and FNAL, 
and recorded hadronic showers, electron showers and muons at
different energies and incident angles.	
\end{abstract}

\section{Introduction}

\begin{wrapfigure}{r}{0.5\columnwidth}
  \centerline{\includegraphics[width=0.45\columnwidth]{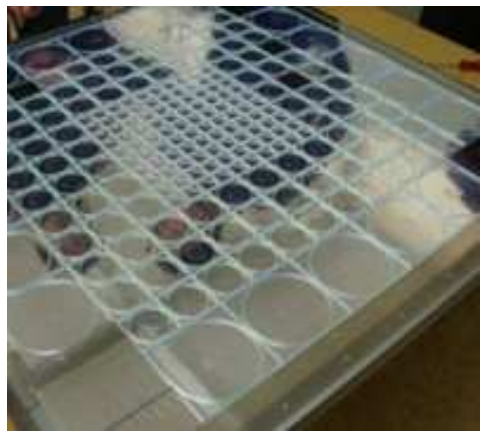}}
  \caption{Picture of one AHCAL active layer}
  \label{Fig:ahc}
\end{wrapfigure}

One fine granular AHCAL prototype layer is shown in figure~\ref{Fig:ahc}. Thanks to the high granularity, the development of hadronic showers can be studied in great detail. The response of the AHCAL to hadronic showers are investigated, both longitudinal and transverse shower profiles in data, and comparisons with the available \textsc{GEANT4} models were already discussed in CALICE analysis note~\cite{calice_had_note}.

Monte Carlo simulations, performed using different models, and including detector and physics effects, are compared with the data. In addition, the mean shower radius is studied as well. The studies on the effect of the shower leakage on the reconstructed energy and energy resolution have been performed too.

\section{Shower starting point}

Hadronic shower development is based largely on nuclear interactions, and therefore hadrons can travel a significant distance before a shower develops. The average distance a hadron travels before a nuclear interaction occurs, which typically starts the cascade of the shower, is called nuclear interaction length $\lambda_{int}$.

\begin{figure}[ht]
  \begin{center}
  \includegraphics[width=0.4\textwidth]{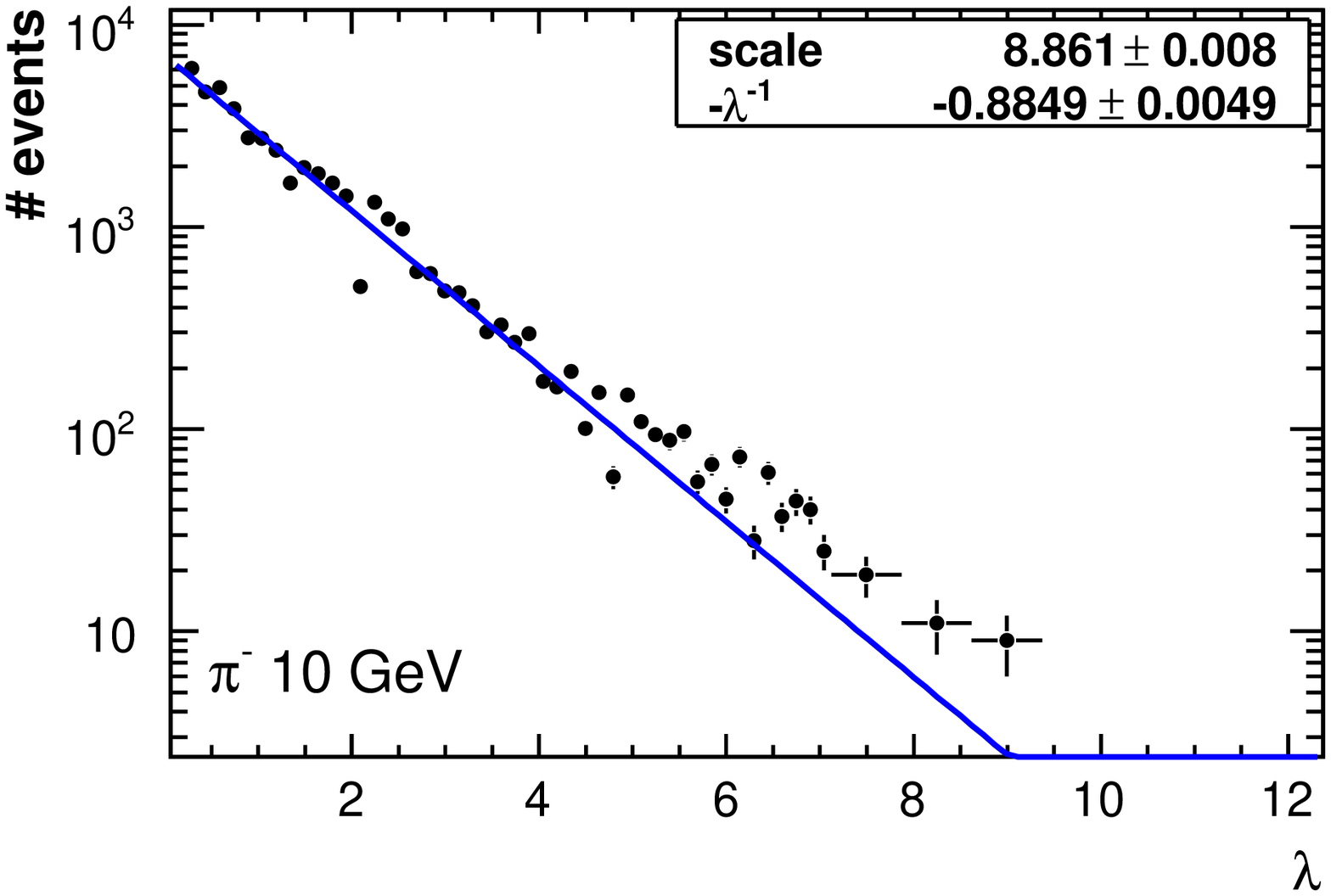}
  \hspace{1cm}
  \includegraphics[width=0.4\textwidth]{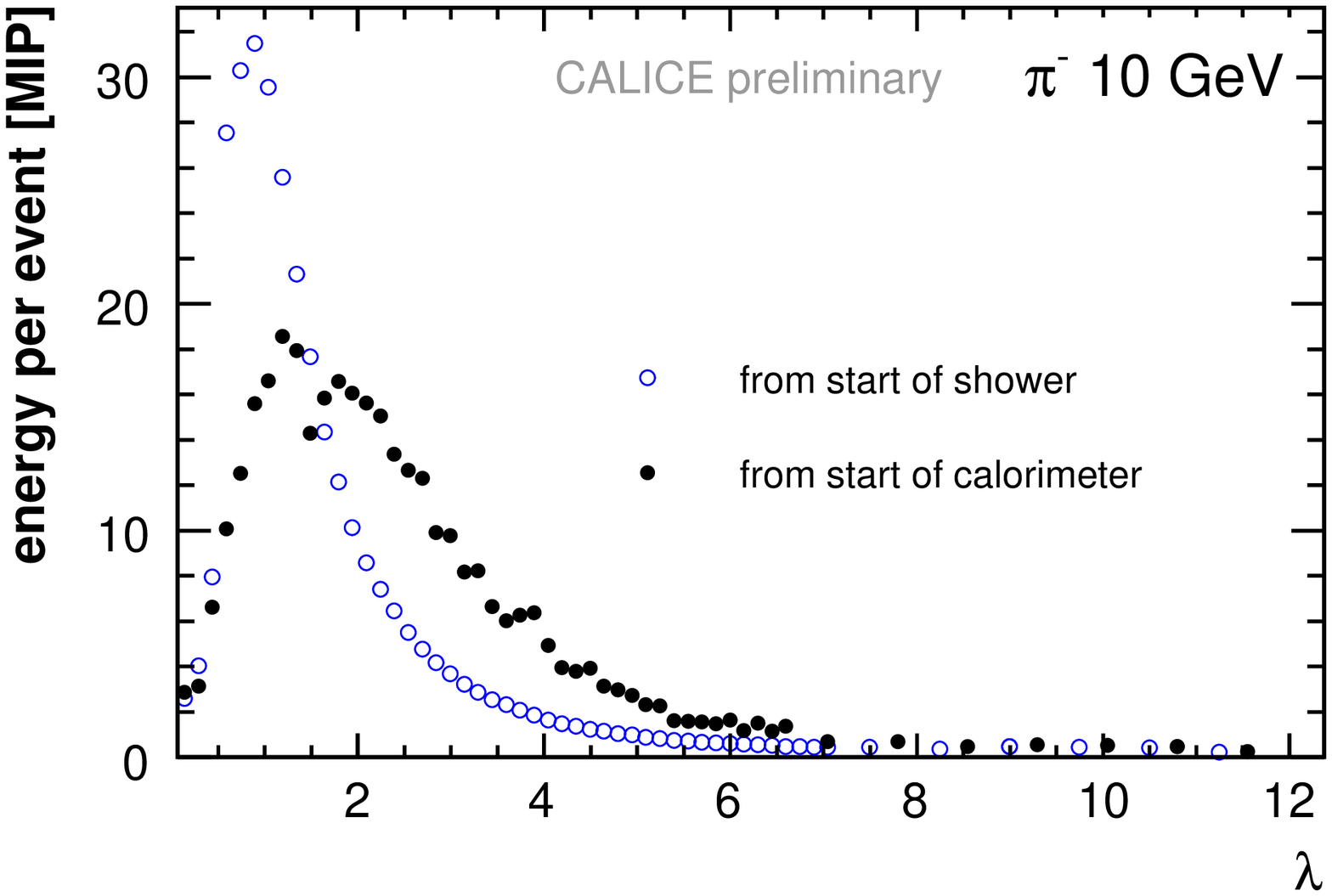}
  \end{center}
  \caption{Position of shower starting point in the calorimeter as a function of the number of interaction lengths (left). Longitudinal shower profile in the calorimeter determined before and after the shift, event by event, to the shower starting point, for 10~GeV pion showers (right).}
  \label{Fig:interaction_length}
\end{figure}

From AHCAL prototype test beam data, the shower starting point as a function of the number of interaction length is studied, the result is shown in the left plot of figure~\ref{Fig:interaction_length}. The decreasing exponential slope agrees with expectations.
When the shower starting point for each event is known, it is possible to shift the single event profile to the same point. The right plot in figure~\ref{Fig:interaction_length} shows the longitudinal profile for 10~GeV pion showers. The open symbols indicate that the longitudinal shower profile becomes much shorter after correcting event-by-event for the variation of the shower starting point.

The measurement of the position of the first hadron interaction allows to disentangle the fluctuations of the position where the shower statrts from those intrinsic in the shower process.

\section{Longitudinal shower profiles Data - Monte Carlo Comparison}

During the CALICE CERN~2007 test beam run period, the energy range from 8~GeV up to 80~GeV have been collected.
A set of runs which were recorded without ECAL in front of AHCAL has been chosen to achieve high statistics and to reduce the selection and energy uncertainties when analysing pions in the AHCAL. Furthermore, runs have been selected where the beam hits the detector centrally to achieve a good radial symmetry. The only data set fulfilling all the requirements was taken with a detector configuration of 28.3~degree with respect to the perpendicular to the beam direction.

GEANT4 based simulations with six different physics lists have been generated for comparison. The simulation includes the following detector specific effects:
\begin{itemize}
\item saturation effects in the scintillator (Birks' law)
\item light crosstalk between scintillator tiles
\item non-linearity of the SiPM
\item statistical fluctuation on the pixel scale
\item limited acceptance in time of the detector electronics
\item detector noise from electronics and SiPM
\item variation of the detector response with temperature
\end{itemize}
 
Due to the generation process, the particle beams in the experiment can contain a significant fraction of muons. These pass the detector as minimal ionising particles without generating showers. The events, where a muon was recorded, have been rejected. Only events, where a shower starting point is found, are used for analysis. The simulations have been generated with pure pion beams.
\begin{figure}[ht]
  \begin{center}
    \includegraphics[width=0.44\textwidth]{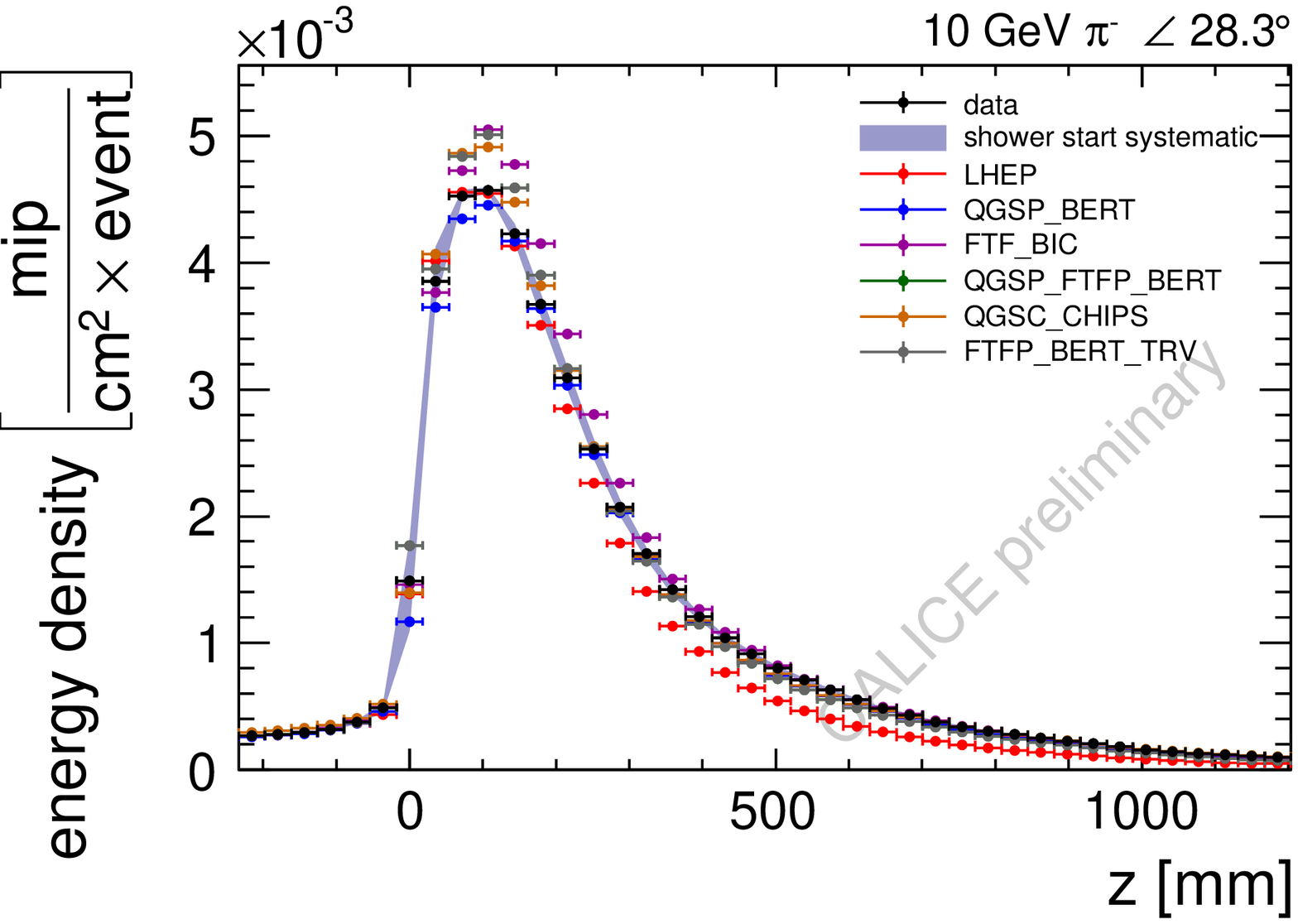}
    \hspace{0.6cm}
    \includegraphics[width=0.44\textwidth]{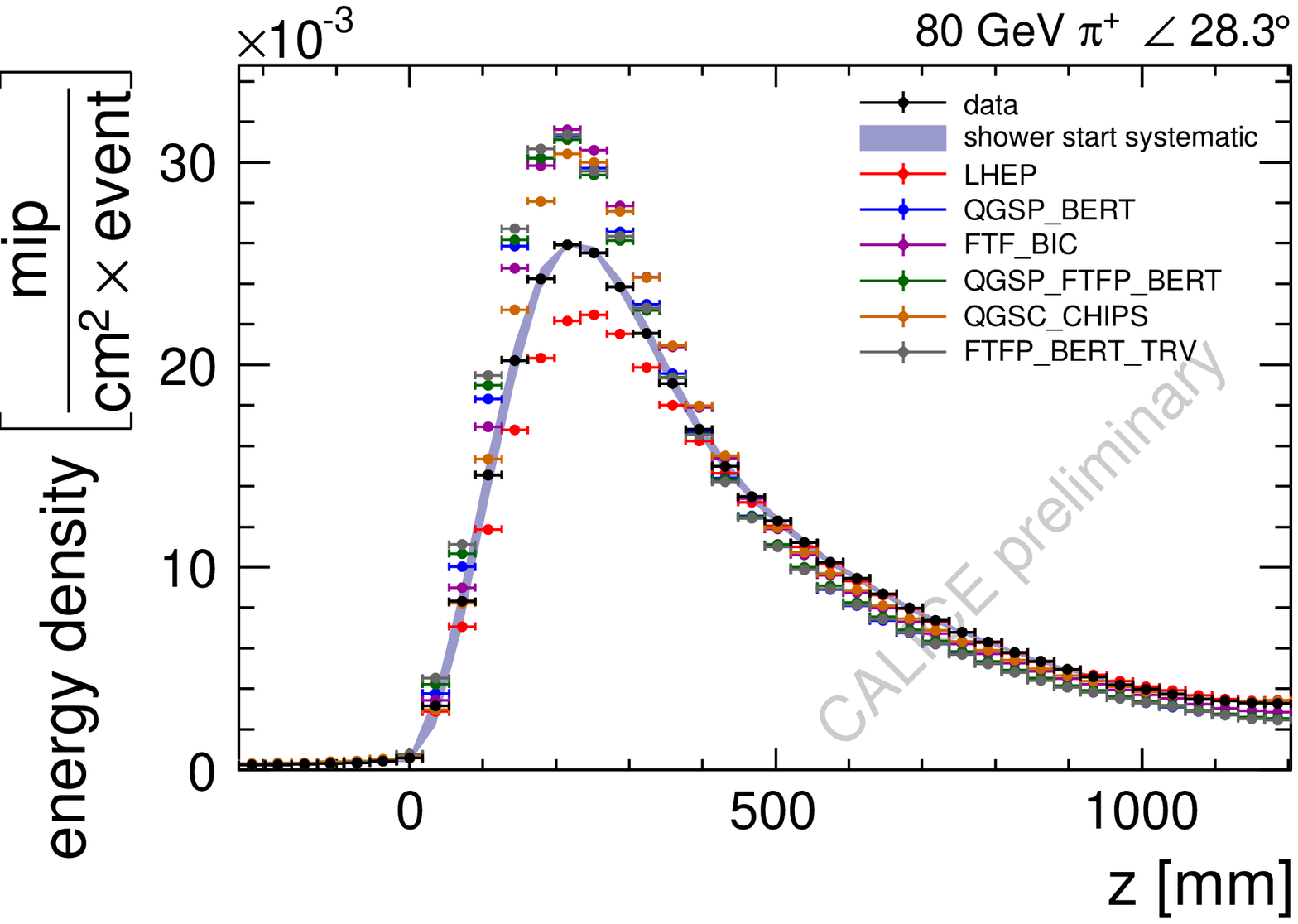}
  \end{center}%
  \caption{Comparison of longitudinal shower profiles from data with selected GEANT4 models for 10~GeV $\pi^-$ (left) and 80~GeV $\pi^+$ (right). The shower starting point has been shifted to the same point event by event.}
  \label{Fig:longitudinal}%
\end{figure}%

The longitudinal shower profiles are shown in figure~\ref{Fig:longitudinal}. The plots also include the predictions of the different simulation models. Peak position and shape of the theory driven models match the longitudinal profile already quite well for 10~GeV. In contrary to this, the parametrised model LHEP cannot describe the longitudinal shape. The radial profile includes some additional discrepancies for QGSC\_CHIPS, which shows a too wide radial distribution. An exact definition of the systematic measurement errors is necessary to select a favourite from the remaining models.

In the case of 80~GeV pions the results is not as clear as for the low energy case. Measurement shows significant differences to all simulation models. This is partly due to artifacts in the measurement that are not reproduced in the digitisation of the simulation. The size of these discrepancies can be seen in figure~\ref{Fig:longitudinal} where the shower max is deformed and the jumps in the tail are not always coherent between measurement and simulation. Nevertheless, these effects seem not to be large enough to explain the discrepancy in the peak shape of the theory driven models and the measurement. These models seem to predict more compact showers, too. LHEP gives again quite different predictions with respect to the other models and the discrepancies to data have opposite sign.~\cite{Beni}

\section{Transverse shower Profiles Data - Monte Carlo Comparison}
\subsection{Transverse Profiles }

\begin{figure}[ht]
  \begin{center}
  \includegraphics[width=0.4\textwidth]{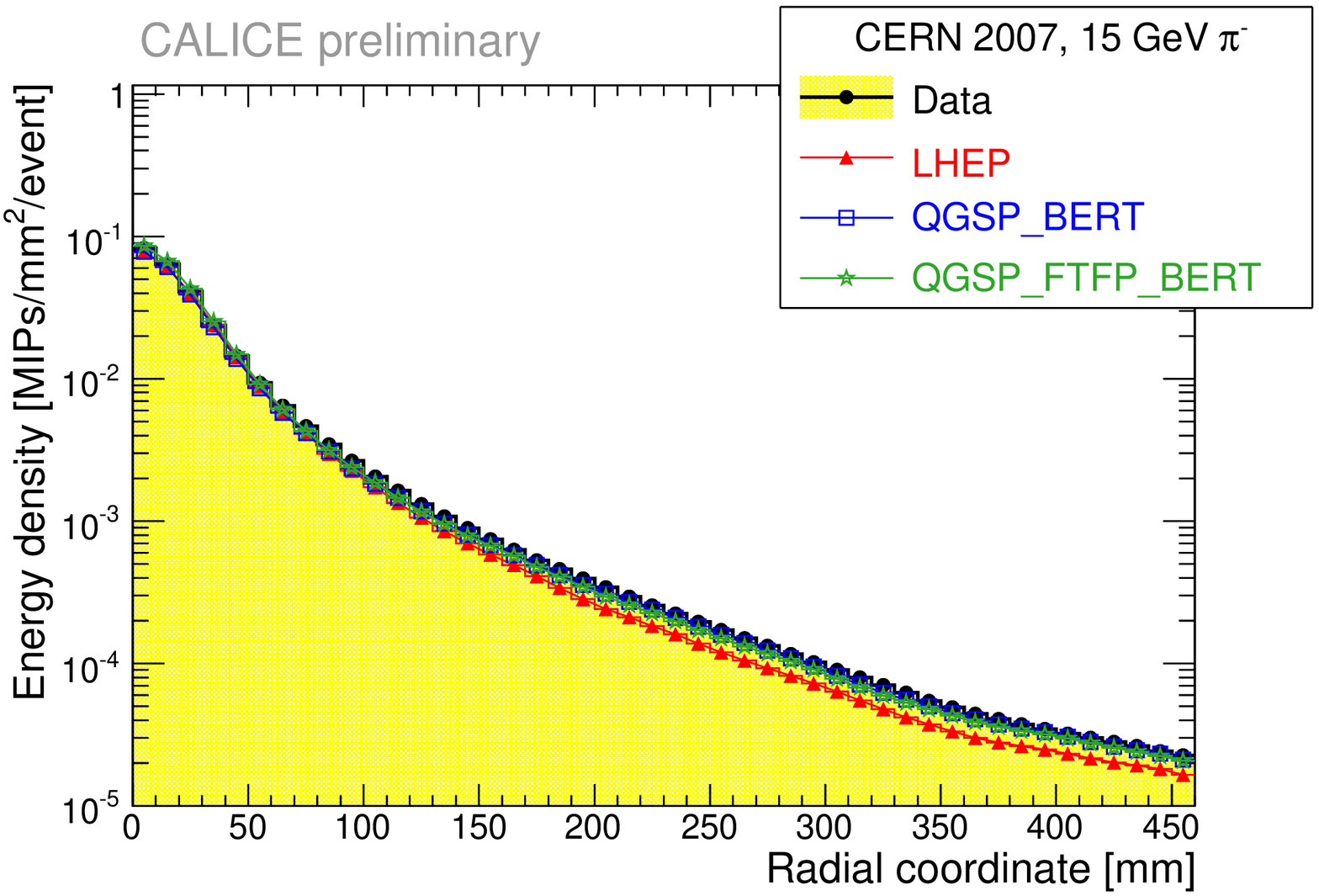}
  \hspace{0.6cm}
  \includegraphics[width=0.4\textwidth]{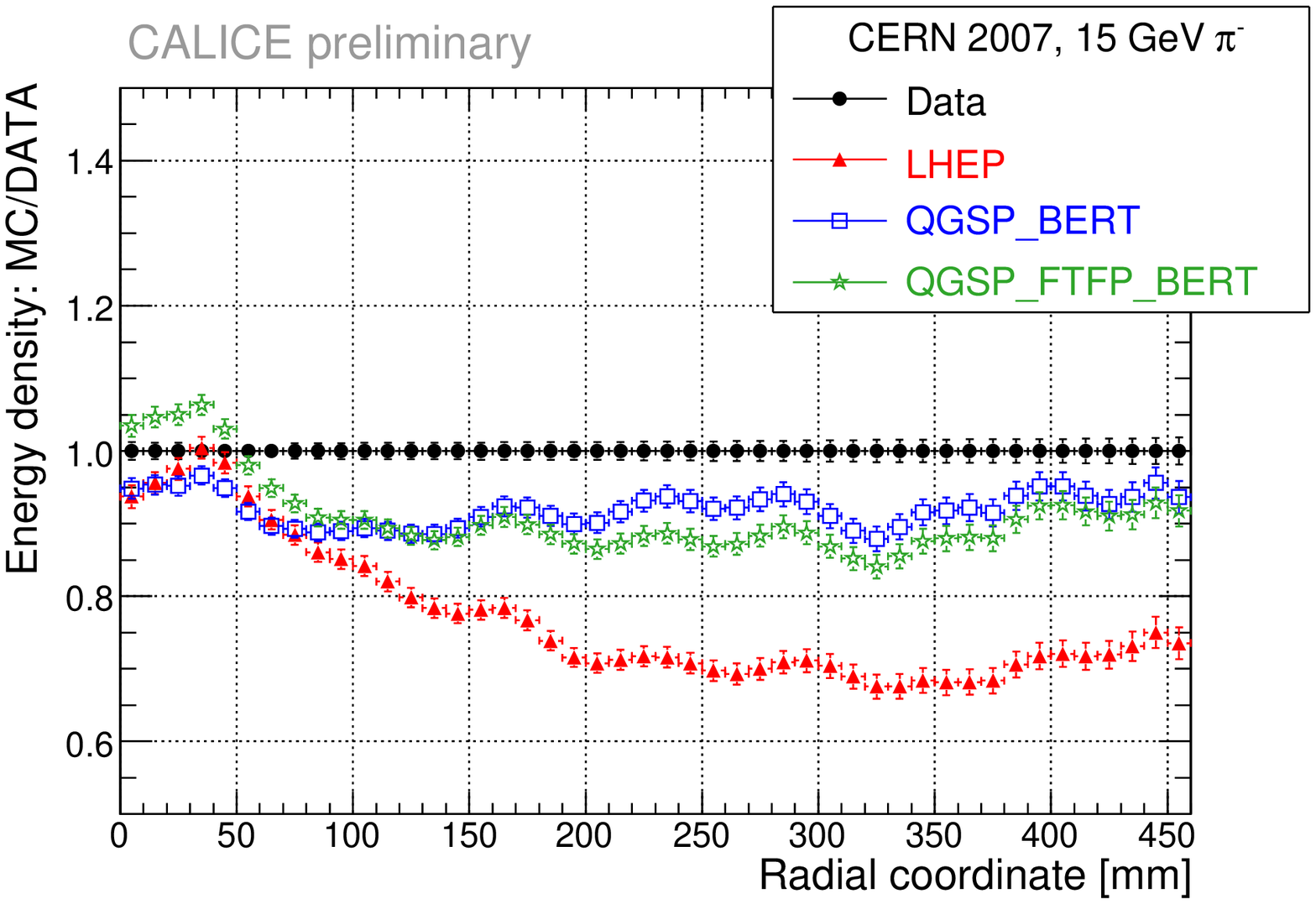}

  \includegraphics[width=0.4\textwidth]{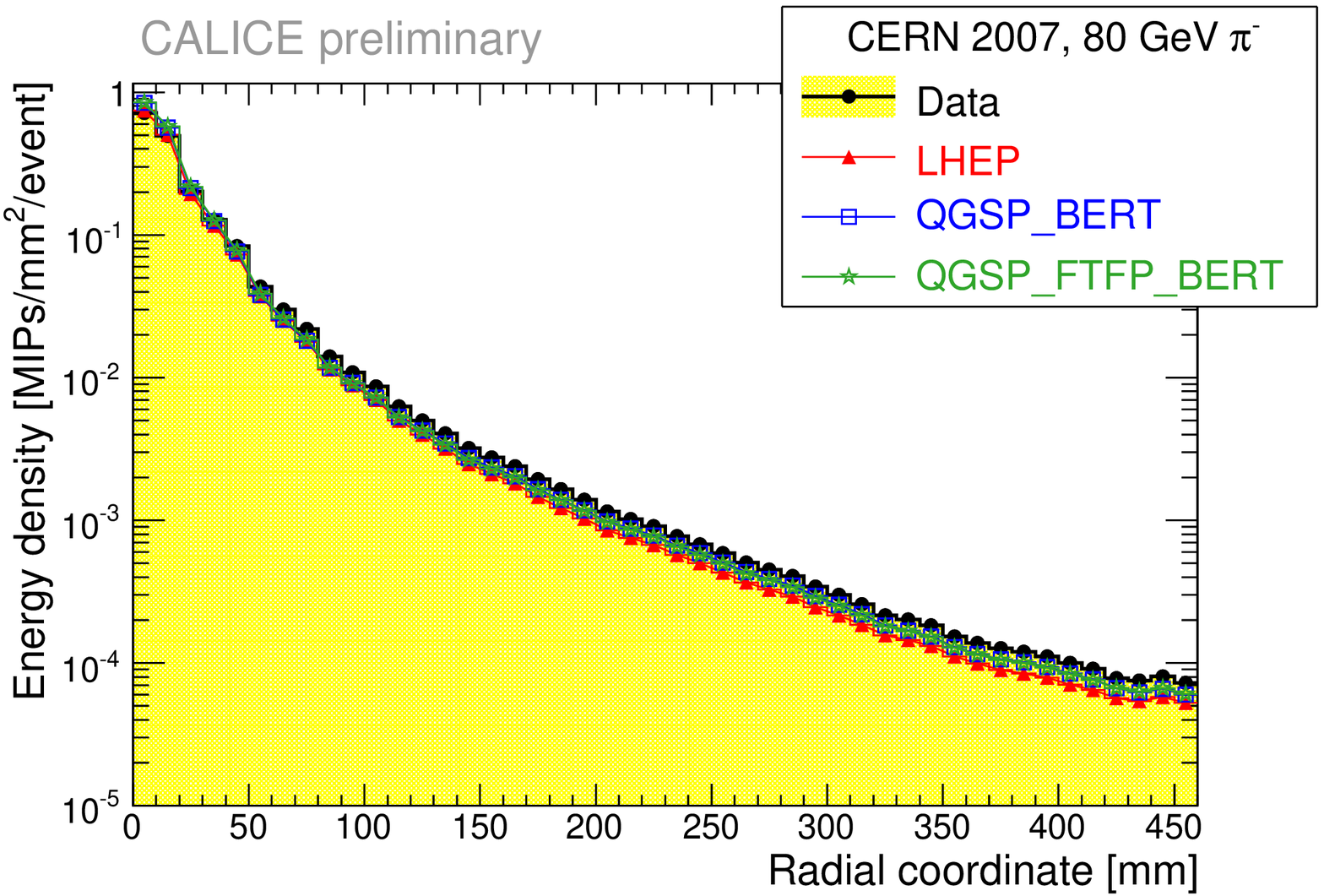}
  \hspace{0.6cm}
  \includegraphics[width=0.4\textwidth]{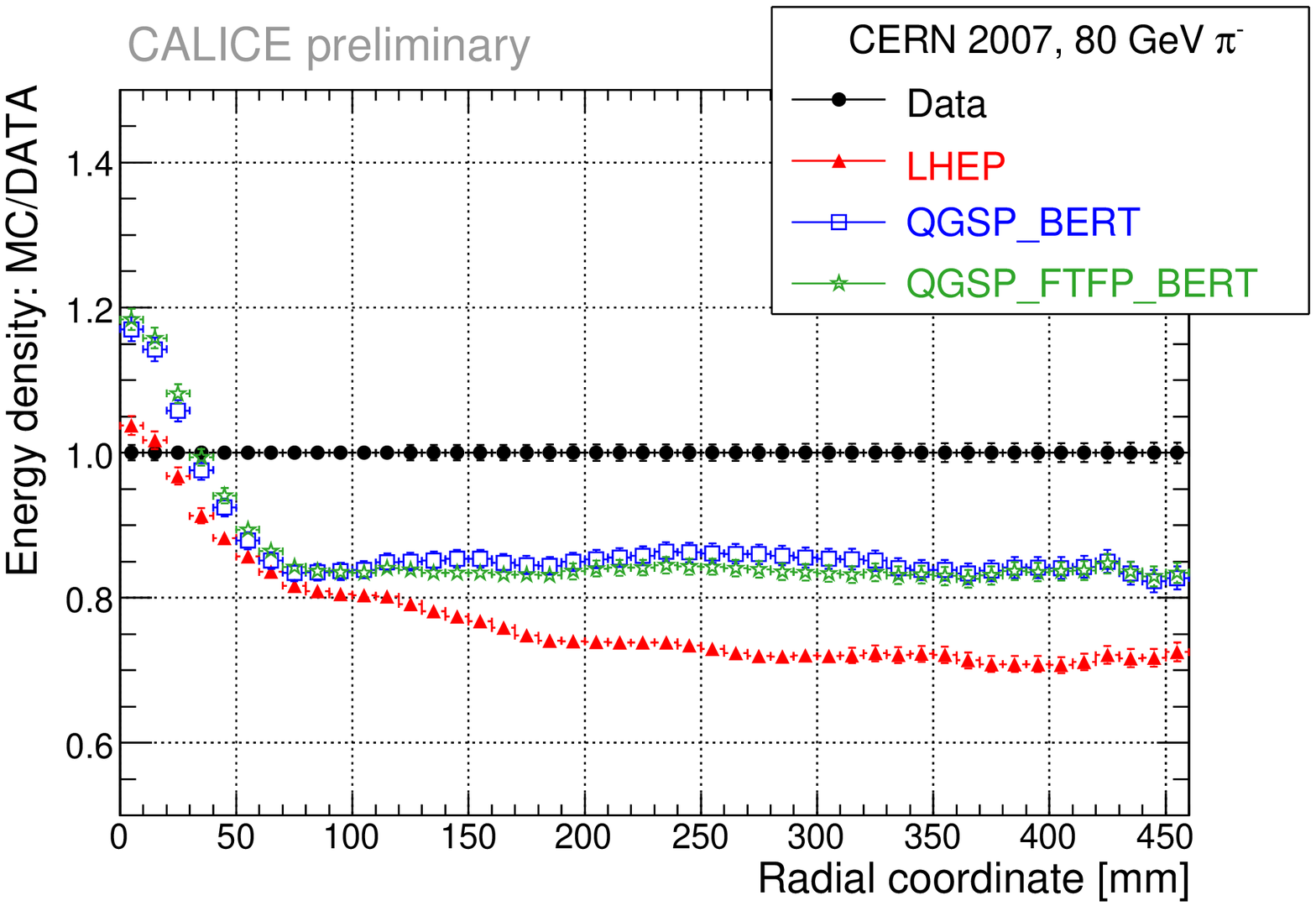}

  \caption{Comparison of the transverse profiles from data with selected GEANT4 models (left) and corresponding data/MC rations (right) for both 15~GeV~$\pi^-$ and 80~GeV~$\pi^-$. \label{Fig:Tran_profile}}%
  \end{center}%
\end{figure}

The comparison of the AHCAL test beam transverse shower profiles with the selected GEANT4 models has also been studied. Figure~\ref{Fig:Tran_profile} shows the transverse pion shower profiles for 15~GeV and 80~GeV. The corresponding ratios are also shown. 
The different GEANT4 models show two distinct behaviours in the shower core, where most of the models are above the data, and in the tails, where almost all are below data. More detail studies on the transverse shower profiles can be found in CALICE analysis note~\cite{calice_had_note_011e,Angela} 

\subsection{Mean shower radius}

\begin{figure}[ht]
  \begin{center}
    \includegraphics[width=0.3\textwidth]{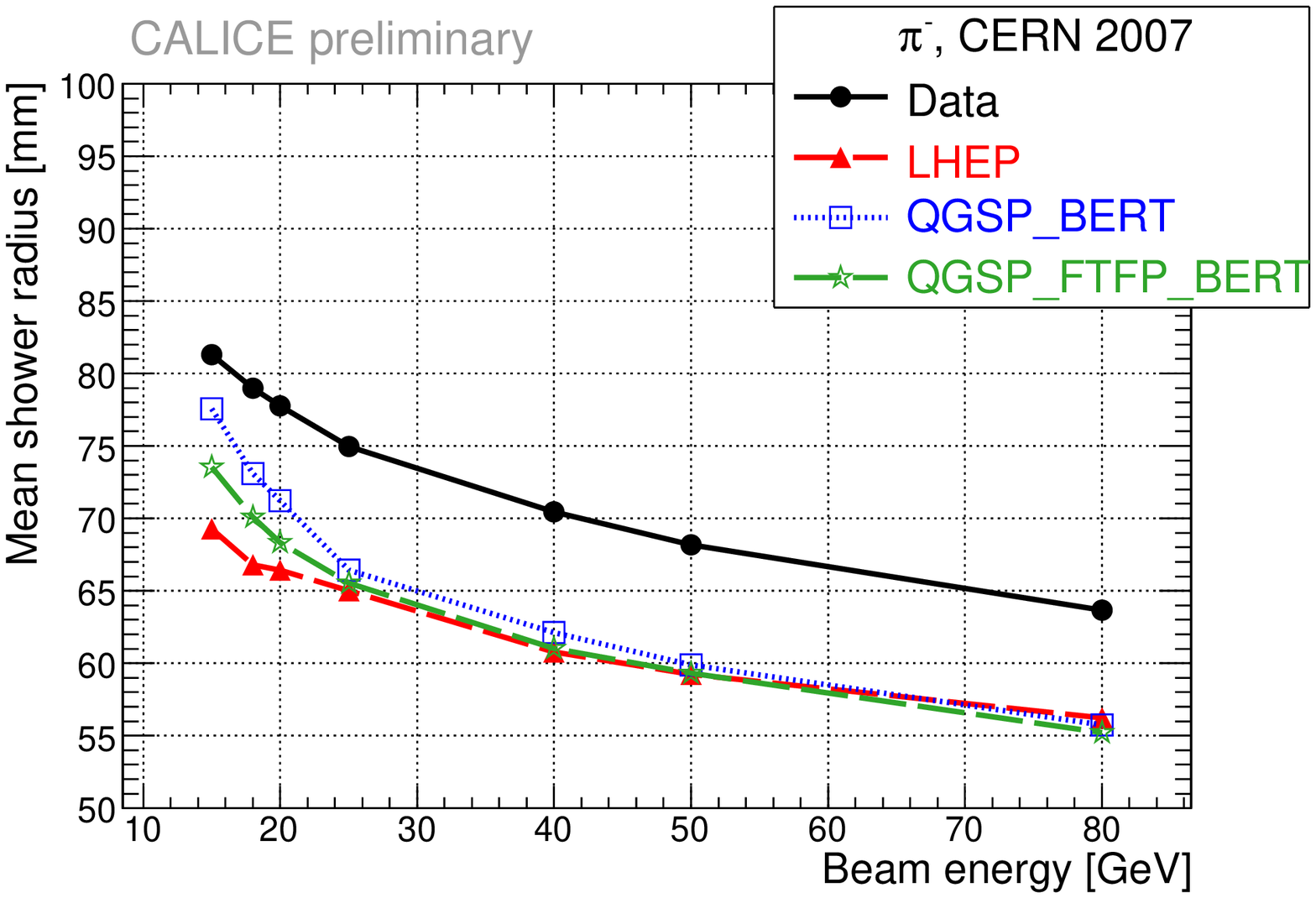}
    \includegraphics[width=0.3\textwidth]{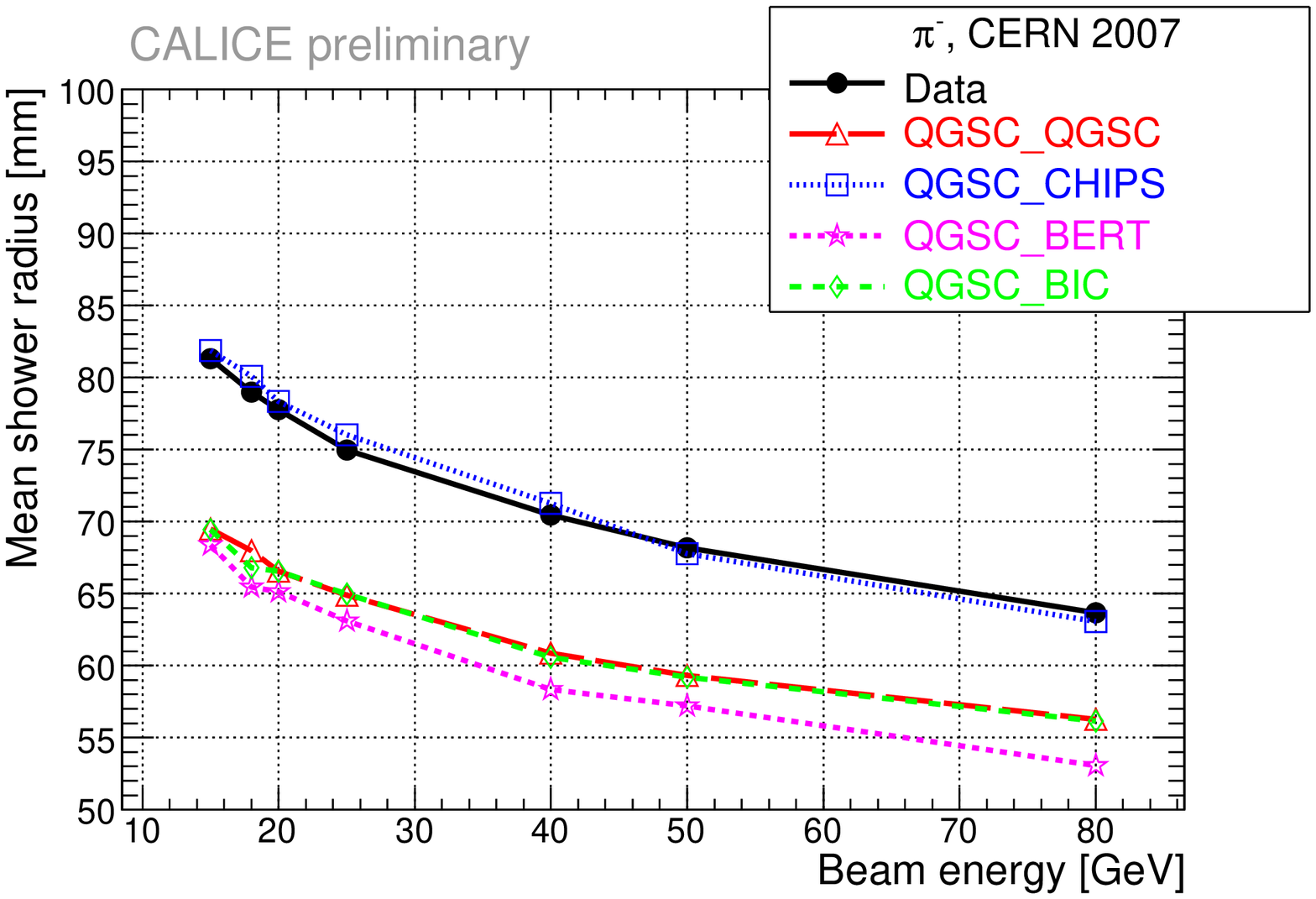}
    \includegraphics[width=0.3\textwidth]{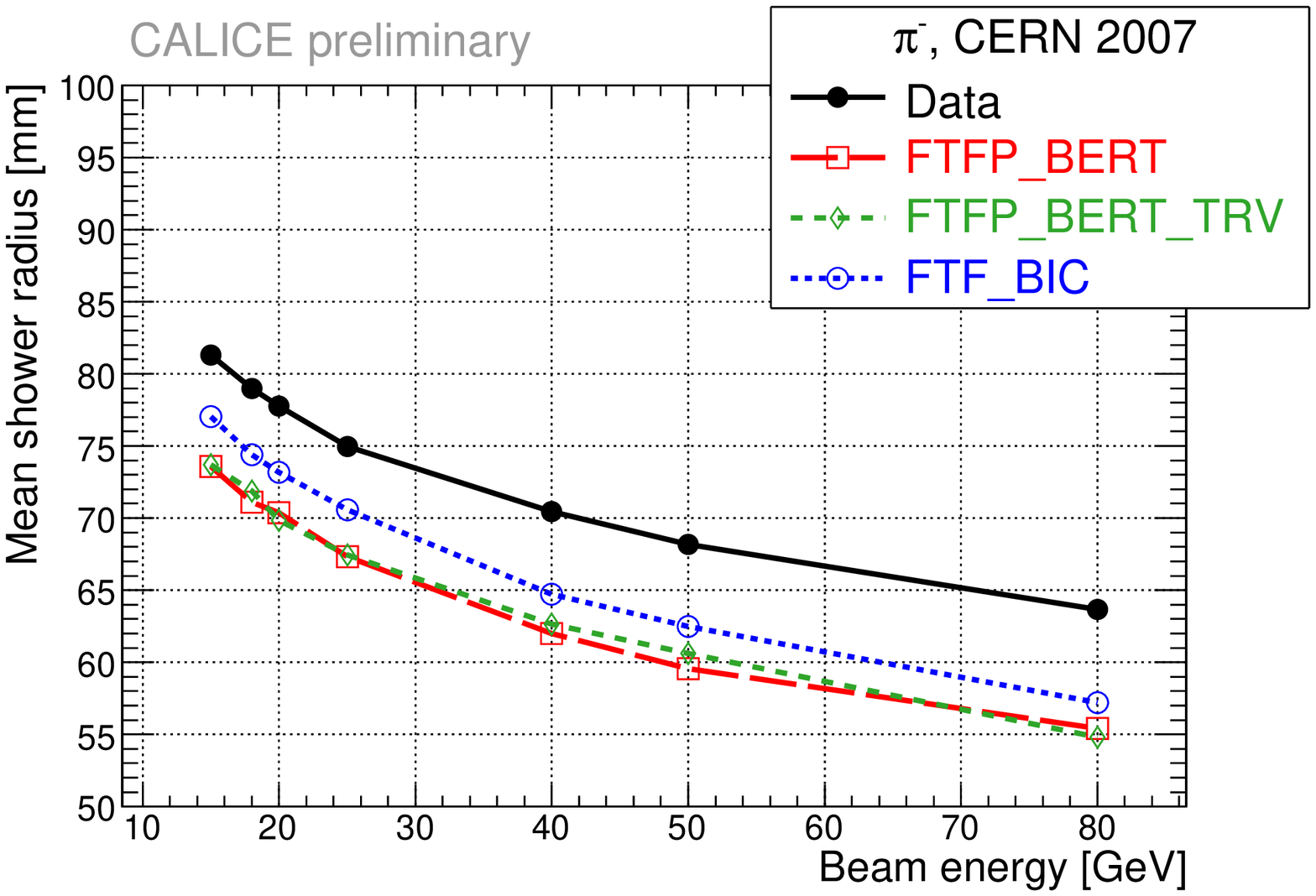}
  \end{center}%
  \caption{Comparison of the mean shower radius measured in data and in Monte Carlo.}
  \label{Fig:showerRadius_allEnergies}
\end{figure}

The mean shower radius are extracted for all analysed energy points from 15~GeV to 80~GeV pions, based on a Gaussian fit in the region containing $90\%$ of the events, and presented in figure~\ref{Fig:showerRadius_allEnergies}.  
With increasing energy, the mean shower radius slowly decreases , i.e. the shower becomes more compact. In general, the selected GEANT4 models have a $\sim15\%$ smaller shower radius than the experimental data. 

\section{Differential Longitudinal Profiles}

\begin{figure}[ht]
  \begin{center}
    \includegraphics[width=0.4\textwidth]{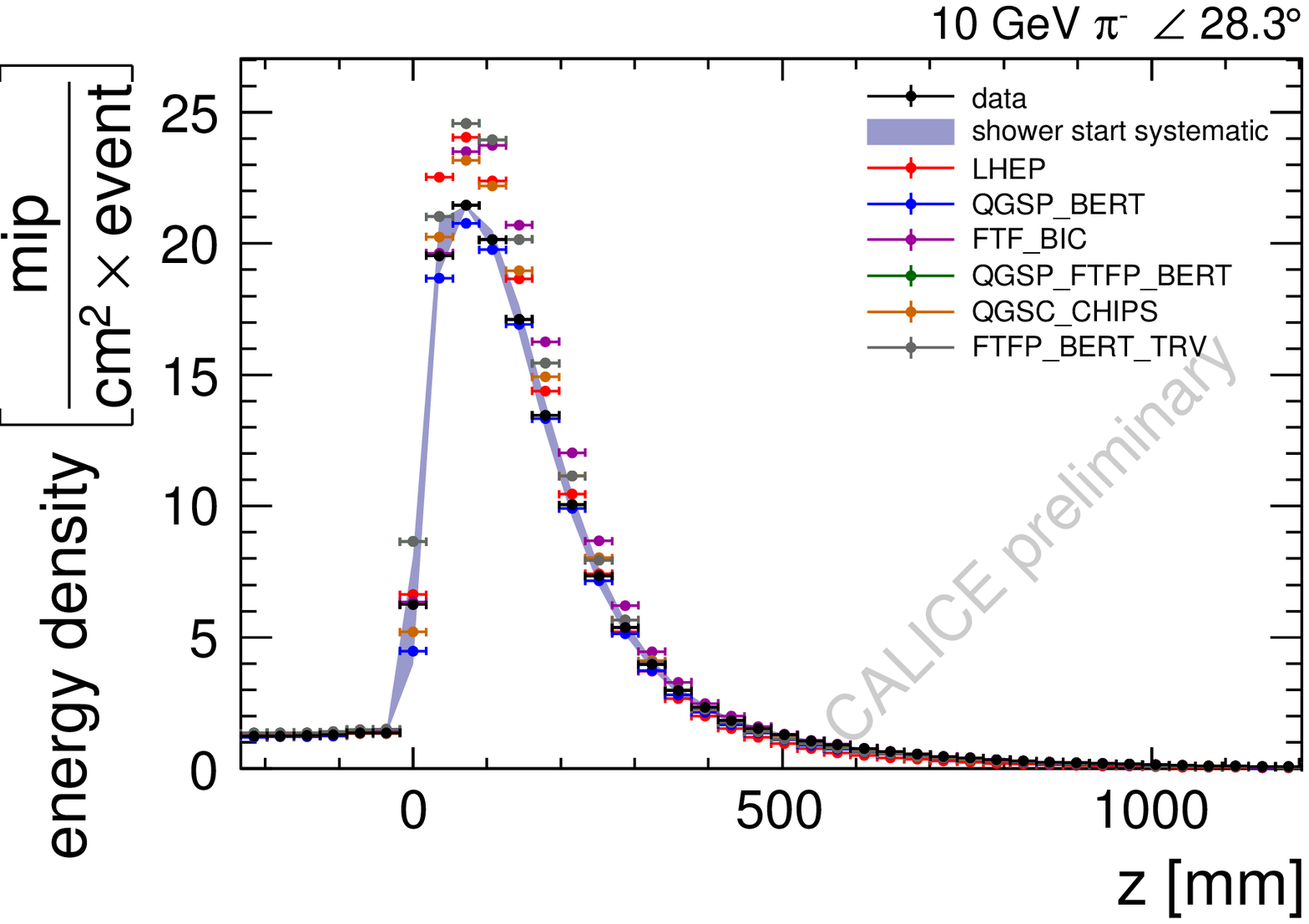}
    \hspace{0.6cm}
    \includegraphics[width=0.4\textwidth]{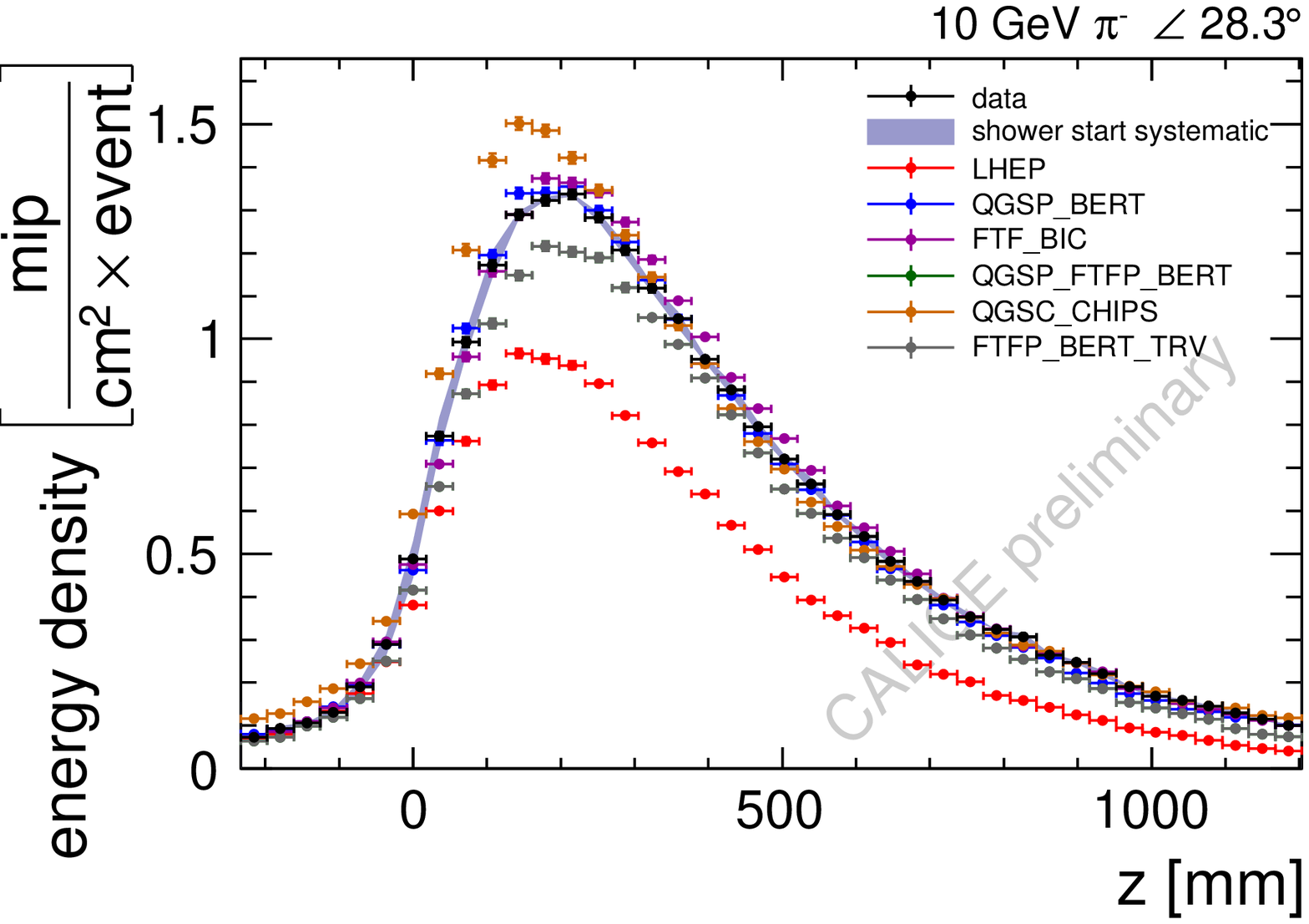} \\
    
    \includegraphics[width=0.4\textwidth]{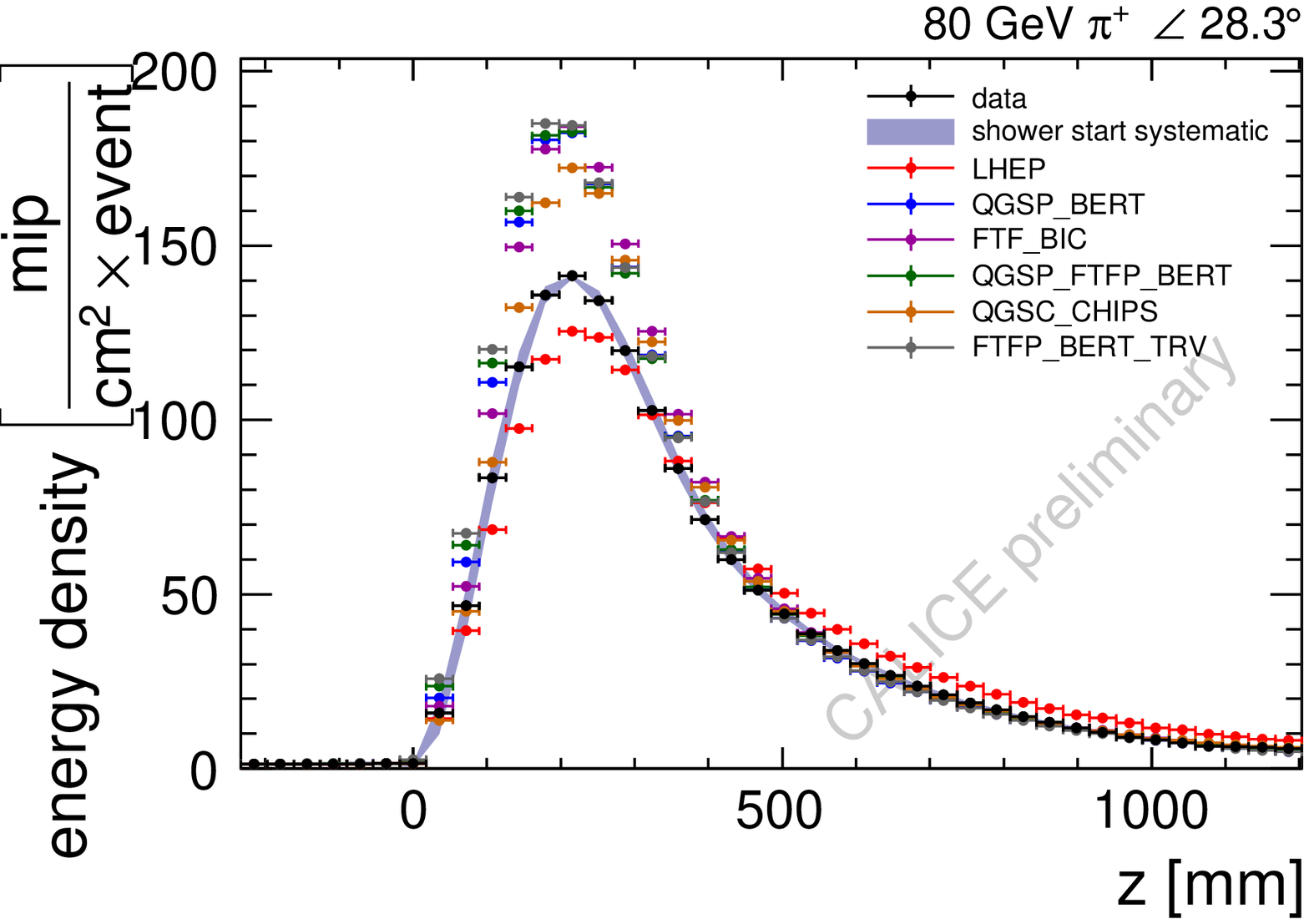}
    \hspace{0.6cm}
    \includegraphics[width=0.4\textwidth]{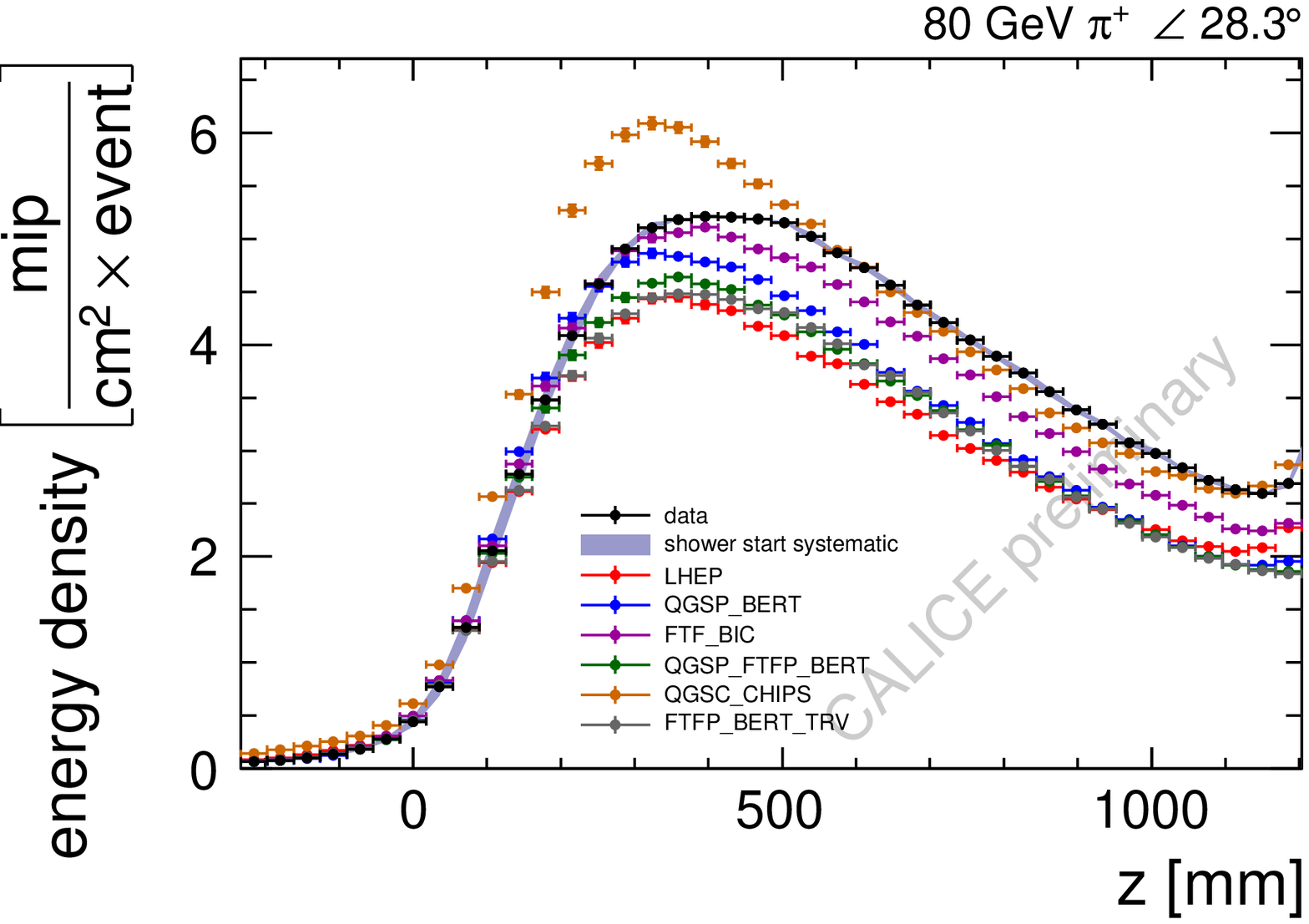}
  \end{center}%
  \caption{Longitudinal profiles from shower starting point for the shower core ( $0\,\mathrm{mm}\le r < 60\,\mathrm{mm}$) left, and the shower periphery ($180\,\mathrm{mm}\le r < 240\,\mathrm{mm}$) right.}
  \label{Fig:profileSlices}%
\end{figure}

The high segmentation and lateral granularity of the calorimeter allow to compare not only integrated longitudinal shower profiles but also differential profiles. This gives more possibilities to test the quality of the predictions from simulation. Figure~\ref{Fig:profileSlices} show the longitudinal profiles measured from the shower starting point for different ranges of the radial distance.

At 10~GeV, the longitudinal shape is reproduced well for all simulation models in the shower core, but several models show significant differences for larger radii.

The sensitivity of this comparison becomes more obvious in the case of the 80~GeV data. Here the picture changes more dramatically between the different radial slices. The integrated longitudinal profile (figure~\ref{Fig:longitudinal}) is strongly dominated by the development in the shower core where all models, except LHEP, give similar results. But, the predictions for the larger radial positions differ significantly and can be used to understand where the different approaches describe the measurements more realistically. FTF\_BIC is the closet model to data in the periphery of the shower.

\section{Shower leakage}
\begin{figure}[ht]
  \begin{center}
    \includegraphics[width=0.3\textwidth]{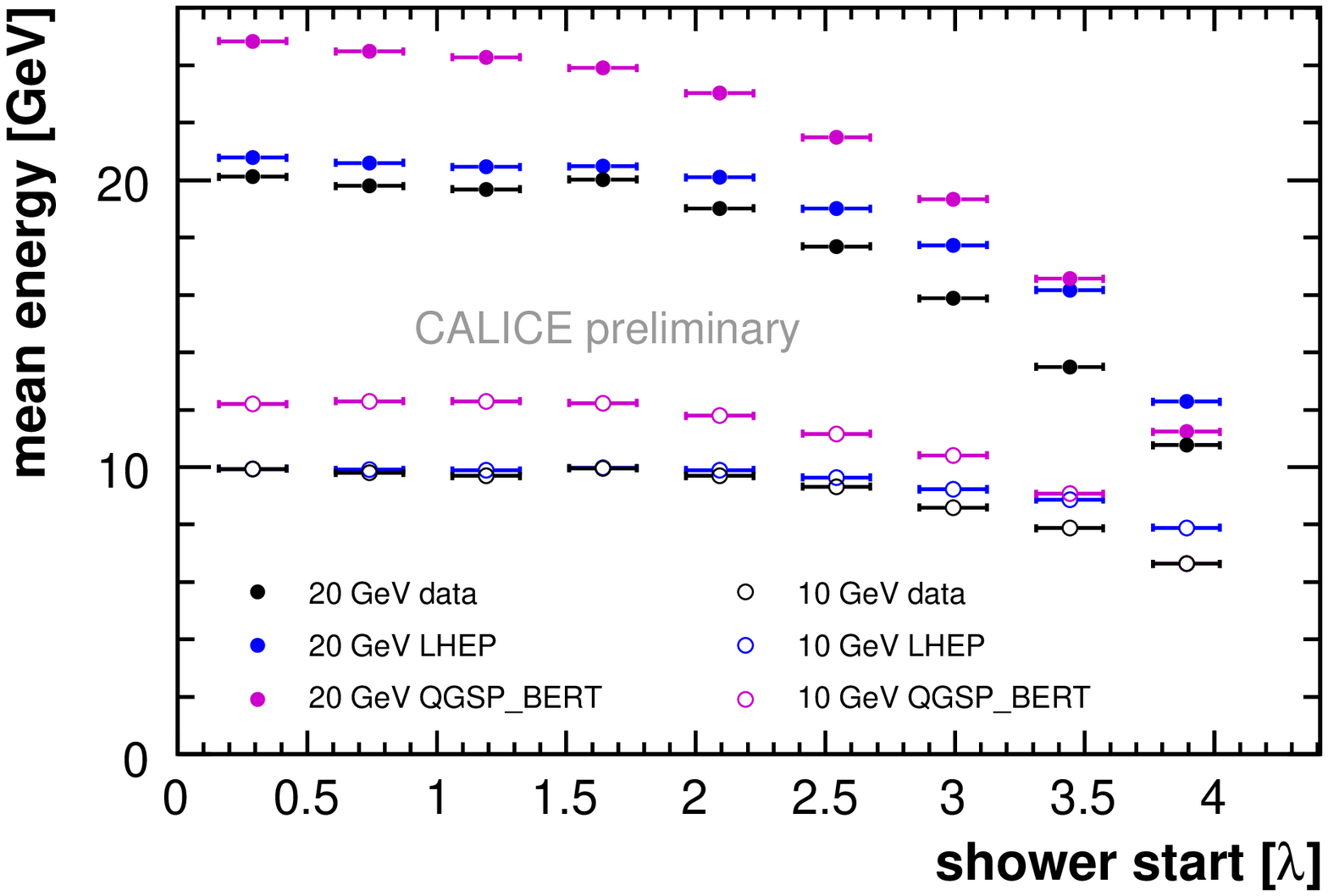}
    \includegraphics[width=0.3\textwidth]{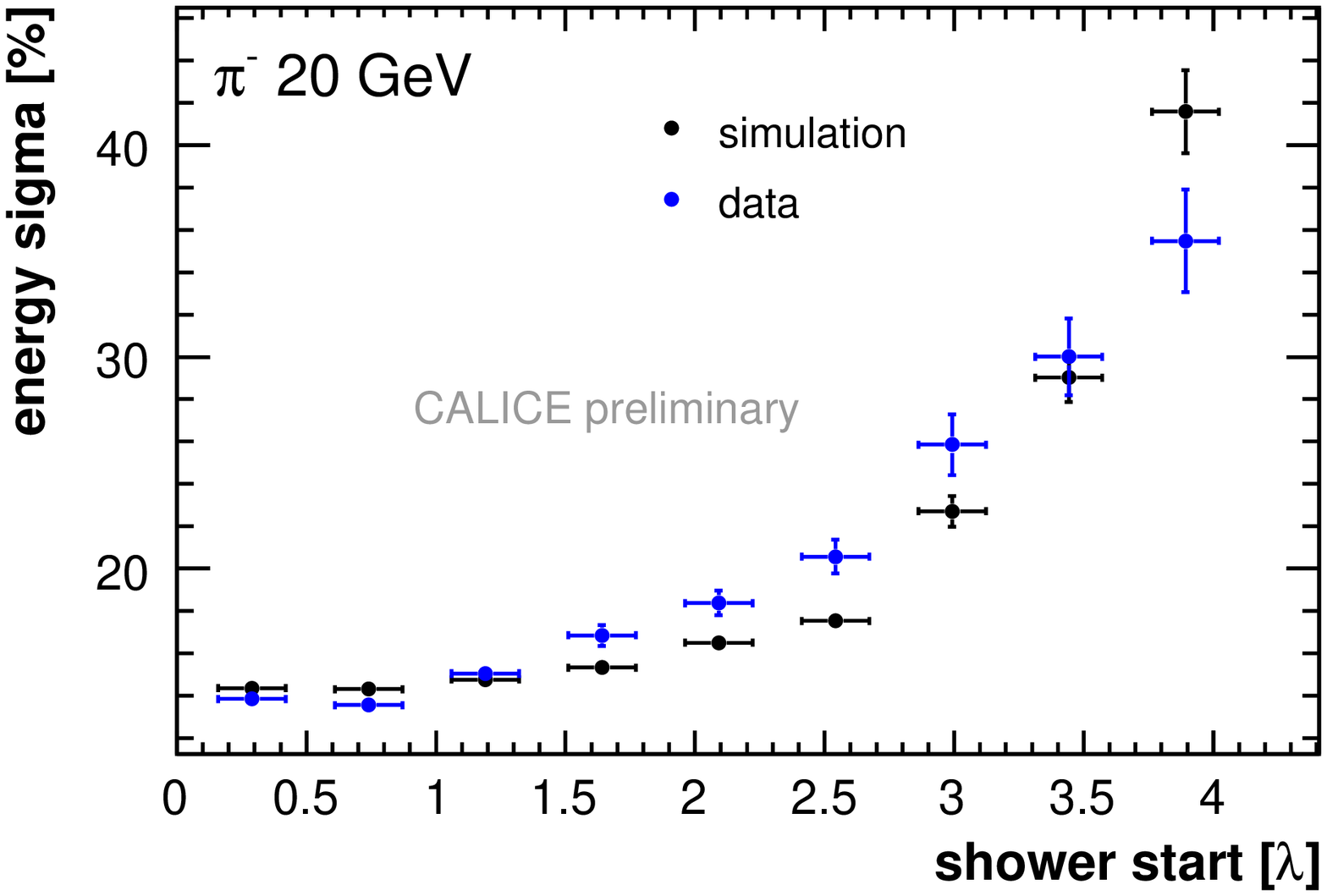}
    \includegraphics[width=0.3\textwidth]{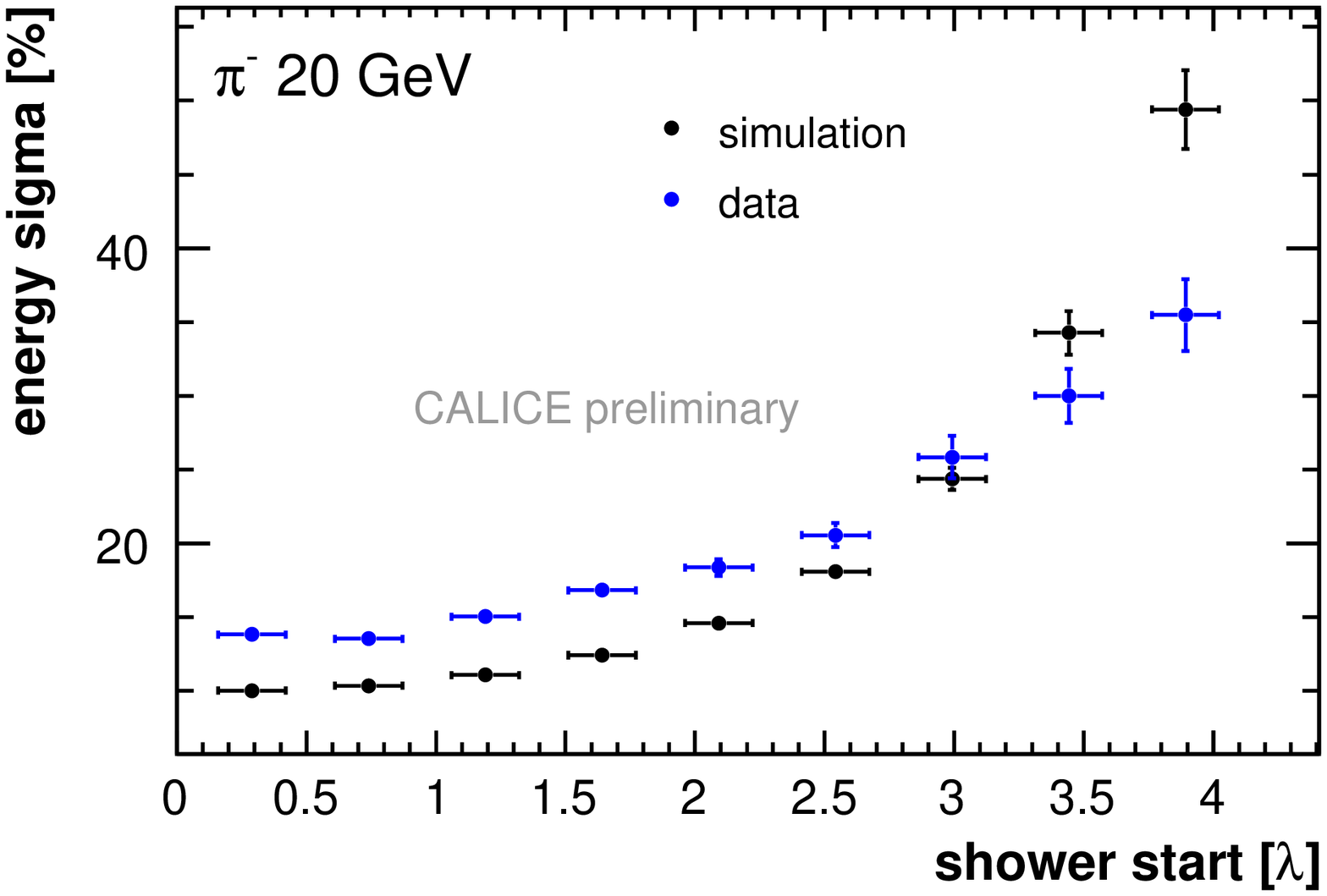}
  \end{center}
  \caption{Comparison of reconstructed energy between data and Monte Carlo (left), comparison of the resolution with QGSP\_BERT (middle) and LHEP (right)}
  \label{Fig:compare_signalVstart}
\end{figure}

The calorimeter response to 10~GeV and 20~GeV poin is presented by the left plot in figure~\ref{Fig:compare_signalVstart} as a function of the shower starting point. The effect of shower leakage is visible for shower initiated after $2~\lambda_{int}$. The energy lost by leakage deteriorates the calorimeter resolution seen in the right two plots in figure~\ref{Fig:compare_signalVstart}.

LHEP shows a slower degradation of the resolution than data (right). QGSP\_BERT predicts a better resolution for early shower but degenerates faster than data for later showers (middle). 
A method to correct event-wise for shower leakage is under study.

\section{Summary}
Hadronic shower development suffers from much larger fluctuations than  electromagnetic ones. In particular the shower starting point varies over a much larger range in the calorimeter depth from event to event. The determination of the shower starting point is crucial to study hadronic shower shapes in detail and to address the problem of shower leakage.
Thanks to the high granularity of the AHCAL prototype, 
The definition of shower starting point is based on topological resolution.
A first qualitative comparison of profiles
have been shown for different simulation models and data recorded.
Qualitatively the GEANT4 models describe the poin shower shape, the agreement is at the $15\%$ level. In general, too short and narrow showers are simulated with respect to data, expecially at high energy.


\begin{footnotesize}


\end{footnotesize}


\end{document}